\documentclass[fleqn,10pt]{wlscirep}

\usepackage{caption}
\usepackage{subcaption}
\usepackage{dirtree}
\usepackage{enumitem}
\usepackage[utf8]{inputenc}
\usepackage[T1]{fontenc}

\usepackage{graphicx}
\begin{document}
\title{ChoralSynth: Synthetic Dataset of Choral Singing}

\author[1, *]{Jyoti Narang}
\author[2]{Viviana De La Vega}
\author[1]{Xavier Lizarraga}
\author[3]{Oscar Mayor}
\author[3]{Hector Parra}
\author[3]{Jordi Janer}
\author[1]{Xavier Serra}
\affil[1]{Universitat Pompeu Fabra, Music Technology Group, Barcelona, Spain}
\affil[2]{Escola Superior de Musica de Catalunya, Barcelona, Spain}
\affil[3]{Voctro Labs, Barcelona, Spain}
\affil[*]{jyoti.narang@upf.edu}



\begin{abstract}
    Choral singing, a widely practiced form of ensemble singing, lacks comprehensive datasets in the realm of Music Information Retrieval (MIR) research, due to challenges arising from the requirement to curate multitrack recordings. To address this, we devised a novel methodology, leveraging state-of-the-art synthesizers to create and curate quality renditions. The scores were sourced from Choral Public Domain Library(CPDL). This work is done in collaboration with a diverse team of musicians, software engineers and researchers. The resulting dataset, complete with its associated metadata, and methodology is released as part of this work, opening up new avenues for exploration and advancement in the field of singing voice research.
\end{abstract}

\maketitle

\section{Introduction}
One prominent challenge with choral singing research is the availability of multitrack datasets. Curating a dataset manually for the case of choral singing is quite challenging, since all singers need to be recorded simultaneously, preventing any leakage between the voice sources. Hence, not many datasets are available for the task. Cuesta, H. et al.\cite{helena_cuesta_2018} present a choral singing dataset that consists of  multitrack renditions of Western choral music. It comprises recordings of three songs in the SATB format, each in a different language - Catalan, Spanish, and Latin. 
Other datasets include Dagstuhl Choirset dataset for MIR research on Choral Singing \cite{rosenzweig2020dagstuhl}, Cantoria \cite{HelenaThesis}, ESMUC Choir Dataset \cite{HelenaThesis}. All of these datasets are publicly available, however, the length of all of these datasets combined is just over one hour of duration, with three of them being less than 20 minutes long.

ChoralSynth dataset takes advantage of the state-of-the-art synthesizers to augment the publicly available datasets for choral singing research. This work is done in close collaboration with diverse team of musicians, software engineers and researchers. In our specific case, the synthesizer has been developed in-house at Voctro Labs\footnote{https://www.voctrolabs.com/}. Through a meticulous curation process, we carefully select and evaluate specific synthetic renditions. In collaboration with a musician, we present a dataset, providing a detailed description of the criteria and methodology employed in its construction, along with detailed documentation outlining the reasons for including each piece. The dataset is available on GitHub\footnote{\url{https://github.com/MTG/ChoralSynth}} and Zenodo\footnote{\url{https://doi.org/10.5281/zenodo.10137883}}.
Overall, the dataset contains 20 choral pieces with each of them consisting of the following:\\
\begin{enumerate}[label=\textbullet]
        \item musicXML score with all voice parts
        \item MIDI file with all voices
        \item multitrack recording containing one track per voice 
         \item  beat positions
        \item metadata consisting of  type of voices in the score like soprano, alto, tenor or bass, language and composer       
\end{enumerate}
 Applications of the dataset include MIR tasks like source separation, beat detection, melodic analysis, or chord analysis.
\section{Dataset Description}
An example structure of the released data for each piece is presented in Figure \ref{verbatim:example_dir}. Additionally, Table \ref{tab: List of composers} offers a summary of the list of composers featured in the dataset, along with their respective musical eras. It is to be noted that most of the composers in the list are associated with Church music.\\

\subsection{MusicXML scores}
The original scores are sourced from Choral Public Domain Library - CPDL\footnote{http://www.cpdl.org/}. It is an online library and repository of sheet music and choral scores that are in the public domain. The library primarily focuses on choral music, and its collection includes compositions from various historical periods and composers.  

\subsection{Audio/Voices Data}
The synthesis technology is provided by Voctro Labs\footnote{https://www.voctrolabs.com/} and the system is based on the Semi-Supervised Learning for Singing Synthesis Timbre algorithm~\cite{bonada2021semi}. The input to the synthesis engine is a score with lyrics, with lyrics tied well with each note for correct rendition. The training data used was a proprietary dataset recorded by Voctro Labs for the TROMPA project, which consisted of recordings by professional choir singers from the Cor Francesc Valls (Barcelona) in Catalan,Latin, Spanish English and German.

\subsection{Beat Times}
The file with beat\_times, refer to the absolute timing of the beats in the score. This value is used to synchronize the audio to the score cursor in the Cantamus  application \footnote{https://cantamus.app/}, allowing to synchronize expressive performances (original choir recordings) that are  tempo -varying. For rendered pieces, these values are automatically generated based on the tempo changes in the score. 
\subsection{Data Statistics}
\begin{figure}
    \centering
    \includegraphics[width=\linewidth]
    {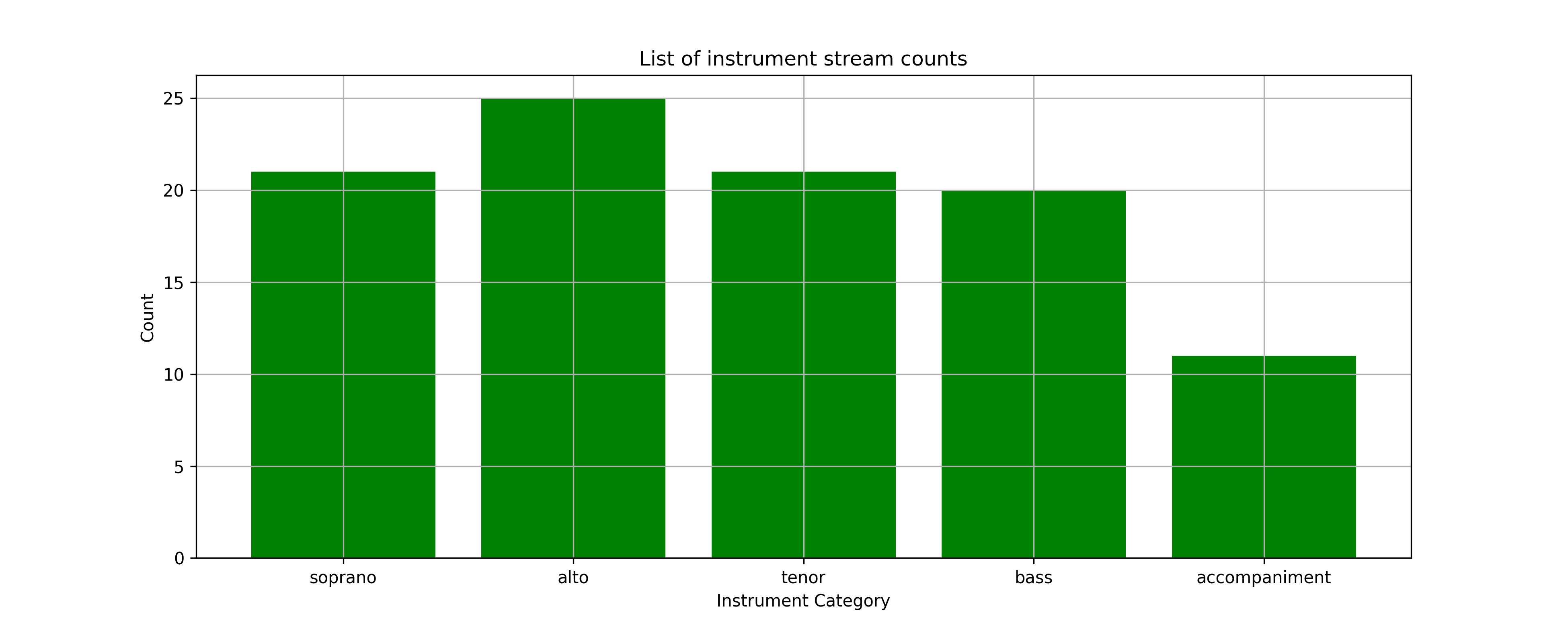} 
    \caption{Barplot of individual streams}
    \label{fig:Data Statistics}
\end{figure}

In Figure \ref{fig:Data Statistics}, we present a statistics of the count of individual streams from all the scores combined. As evident from the figure, all the scores consist of SATB - Soprano, Alto, Tenor and Bass, with alto stream being the most present. 11 scores of the same also have accompaniment streams. 

\begin{figure}
    \begin{tabular}{cc}
        \begin{subfigure}{0.5\linewidth}
            \includegraphics[width=\linewidth]
            {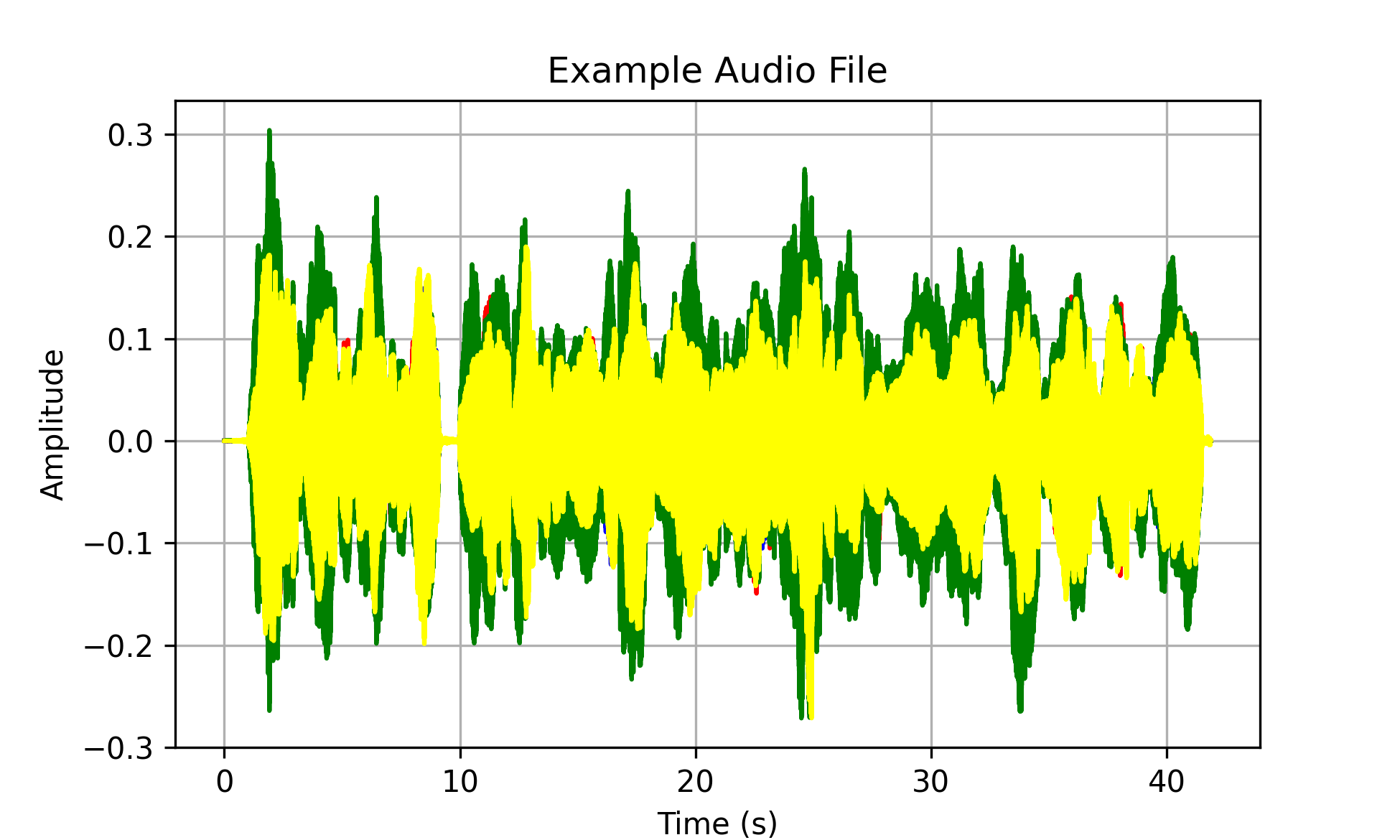} 
        \caption{Example Waveform}
        \label{fig:Example Waveform}
        \end{subfigure}
    &
        \begin{subfigure}{0.5\linewidth}
            \includegraphics[width=\linewidth]
            {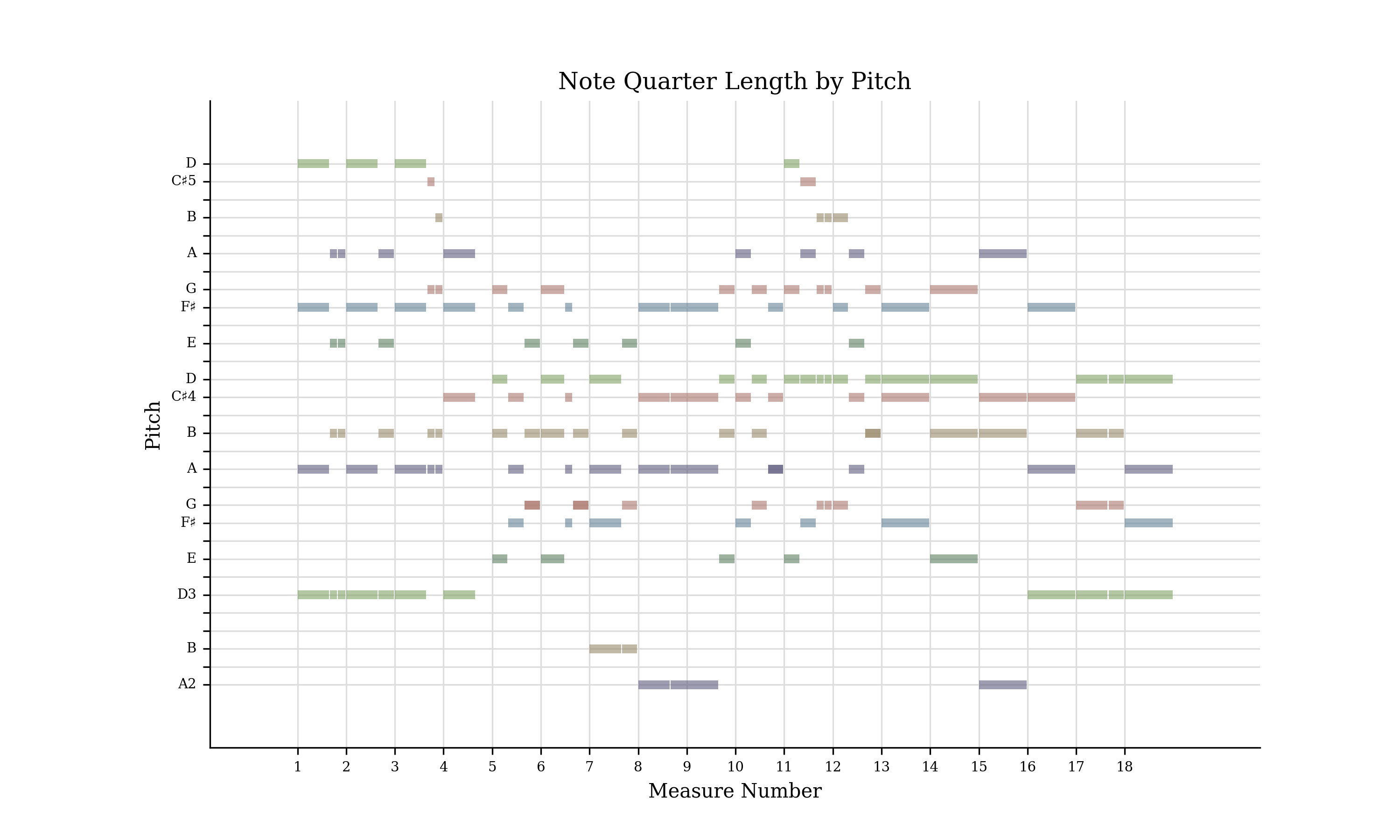} 
            \caption{Example Piano Roll}
            \label{fig:Example Piano Roll}
        \end{subfigure}
    \end{tabular}
    \caption{Example Waveform and MIDI}
    \label{fig:Example Waveform with MIDI}
\end{figure}

In Figure \ref{fig:Example Waveform}, we present the waveform visualization of the streams of synthesized data of the same example as presented in Figure \ref{tab:example_description}, and in Figure \ref{fig:Example Piano Roll}, we present the piano roll of the same waveform, which we extract using  from the musicXML score. Using the musicXML notation, we extract MIDI representation and make them available alongside the other meta data on the GitHub link. 

The overall dataset duration that we release as of now is about 3.8 hours.
Based on language, we currently include 12 songs in English, 7 in Latin and 1 in German.

\begin{figure}
    \centering
    \begin{minipage}{.6\textwidth}
        \begin{verbatim}
          / (ChoralSynth)
          |-- 08_Anima_nostra
          |   |--beat_times.json
          |   |--config.json
          |   |--info.json
          |   |--score.musicxml
          |   |--score.midi
          |   |--voices
          |   |   |--ALTUS.mp3
          |   |   |--BASSUS.mp3
          |   |   |--CANTUS.mp3
          |   |   |--TENOR I.mp3
          |   |   |--TENOR II.mp3
        \end{verbatim}
        \caption{Example Directory Structure of a Folder}
        \label{verbatim:example_dir}
    \end{minipage}
\end{figure}
\section{Methodology}
Our analysis begun with a dataset of 889 synthesized scores, each accompanied by its respective metadata.

\begin{table}[htbp]
    \centering
    \begin{tabular}{|l|l|}
        \hline
        \textbf{Composer} & \textbf{Composition Count} \\
        \hline
        Giovanni Pierluigi da Palestrina (1525-1594) & 1 \\
        \hline
        Leone Leoni (1560-1627) & 1 \\
        \hline
        Thomas Tomkins (1572-1656) & 2   \\
        \hline
        Sigismondo d'India (1582-1629) & 4   \\
        \hline
        Maurice Greene (1696-1755) & 1 \\
        \hline
        Benedetto Re (17th century) & 1  \\
        \hline
        Johann Michael Haydn (1737-1806) & 1  \\
        \hline
        Charles William Hempel (1777-1855) & 1  \\
        \hline
        George Alexander Macfarren (1813-1887) & 1   \\
        \hline
        Henry David Leslie (1822-1896) & 1   \\
        \hline
            Myles Birket Foster (1825-1899) & 1  \\
        \hline
        Walter Cecil Macfarren (1826-1905) & 1 \\
        \hline
        Charles Hubert Hastings Parry (1848-1918) & 1  \\
        \hline
        Roger Petrich (1939-2022) & 3  \\
        \hline
    \end{tabular}
     \caption{Composer and Composition Data}
     \label{tab: List of composers}
\end{table}

\subsection{Criteria for Selection}
Using 889 scores with all its meta-data, the initial step involved the manual design of filtering criteria. This design was based on a careful examination of 276 folders and their corresponding synthesized versions to ensure comprehensive and relevant criteria. Subsequently, the designed criteria were used for automatic filtering of the scores.
We present here the step by step procedure that we followed in the process of selection. A python implementation of the same is presented in the supporting Github repository.

\subsubsection{Automatically Supported Criteria}

First, we add a set of automatic criteria for filtering out scores to be analyzed. We use a python  library music21\cite{cuthbert2010music21} for this part. The automatic criteria includes:
\begin{enumerate}
    \item Filter all folders in which the musicXML scores are not parsable i.e. cannot be read using musicXML reader using .

    \item Filter all folders for which the count of the number of streams in ‘voices’ folder is not the same as the number of streams in the musicXML file read using .
    
    \item Filter out all folders with 'Tr..mp3' as one of the synthetic audio file. The folders with 'Tr..mp3 were discarded because after hearing, it was realized that, the audio always had an instrumental voice instead of singing voice.
    
    \item Filter all the folders for which, one or more voices in the score have no lyrics. Even though some editors avoid to repeat the same lyric for all voices, assuming that the musicians will relate to the first voice to read the corresponding verse, the synthesis however consists of audio files that are instrument generated instead of voice generated for those cases.    
\end{enumerate} 
\subsubsection{Manually supported Criteria}

At the end of automatically supported criteria, we obtain a total of 586 folders. A script for retrieving the folders is included in the supporting GitHub repository. However, this extensive list is impractical to listen to in one go. Hence, we initiate our analysis by dividing the folders into blocks of 20, arranged in alphabetic order. Within each block, the musician first assesses a single audio track, gauging its suitability for further examination. Once a folder is deemed viable during this initial assessment, it becomes a candidate for more detailed manual inspections. The additional manual checks include: 
\begin{enumerate}
    \item If the complete lyrics of the chorale are located exclusively in the first note instead of each corresponding syllable-note pair, the score is discarded, since, in that case, the synthesized voice only outputs vowel "a".
    \item If the chorale has several verses, and the corresponding repeat indicators have been ignored in the .mscz file, the score is discarded because the synthesized voice only occurs for the first verse.
    \item When the chorale has different stanzas and in the transcription of the score, each different stanza appears in different voices, instead of each and every one of them appearing under each of the voices with the corresponding repetition notation, the score is discarded because the synthesized voice of each channel responds to different texts instead of being generated as textually simultaneous channels. It is to be noted that this is not related to canon, a perfectly valid structure for a choral ensemble; this has been differentiated when relevant.
    \item When the notation that marks the separation of a syllable that follows it is a short dash attached to the last letter instead of a long dash equally spaced between both syllables, the synthesized speech presents an error in the pronunciation of the language in question. If at least one of these situations is detected in the visual check of any of the parts, the score is discarded. This is probably due to a characteristic of the editing of the .pdf source.
    \item In Anglo-Saxon languages, when words are written in a contracted way i.e by replacing some syllable or letter with an apostrophe, the synthesis is distorted because the words in question are not taken as the same word or phoneme, but rather a spelling of them is synthesized. The corresponding scores are discarded.
    \item If the voice polyphony corresponding to the last concluding note of the final cadence is absent, the score is discarded since said resolution will not be synthesized and this implied a musical type error. This probably occurs because in ancient chorales, the last notes of the piece occupy the entire value of a measure, which in that case is of a long duration, or "square" in rhythmic terms, and somehow the synthesis does not take that into account.
 
\end{enumerate}
Once the above criteria is fulfilled, we chose the voice that has greatest rhythmic-melodic density or the one that presents the greatest range of register, since they will probably be the ones that can present the greatest distortions. In case, the selected channel meets the compliant tuning and/or linguistic requirements, the folder (score along with all its meta-data) is marked `viable'. At the end of this iteration, the folders that were initially listed as `viable' are manually analyzed again, this time for a complete audio validation. If the final validation is passed, the folder is marked as accepted. Otherwise it is rejected. An example case of analysis is presented in Table \ref{tab:example_description}. An excel file containing the analysis of about 60 songs is attached along with the dataset in the GitHub repository. It is to be noted that only 20 have been accepted, the reasons for rejection are also elucidated. 
\begin{table}[htbp]
    \centering
    \begin{tabular}{c|p{5cm}|c|p{3cm}}
    \hline
    \textbf{Song Name} & \textbf{Song Description} & \textbf{Status} & \textbf{Comments} \\
    \hline
     Anima nostra & Cantus firmus plus 4 (A,TI, TII,B). Text: Ulenberg. Music: GP da Palestrina (transcripted from the original). Tempo: 96 & ACCEPTED &Same note articulated one behind the prior, not almost distinguishable \\
     \hline
      & Cantus Firmus &  & Transcription OK. Pronuncia OK. Cantus vs Tenor I verified, they are aligned. \\
    \hline
    & Altus && Transcription OK. Pronuncia OK. Long note in measures 9 and 10, is being opened as if it were pop, out of style. To take into account. \\
    \hline
    & Tenor I && Transcription OK. Pronuncia OK. \\
    \hline
    & Tenor II && Transcription OK. Pronuncia OK. \\
    \hline
    & Bassus && Transcription OK. Pronuncia OK. \\
    \end{tabular}
    \caption{Example Manual Description of the Selected Songs}
    \label{tab:example_description}
\end{table}

\section{Discussion}

While we've established an alternative approach for analysis that doesn't rely on recording real singers, it's important to emphasize that this setup cannot serve as a direct substitute for live, human singers.

In the analysis of musicXML scores, it was found that the lyrics information is sometimes included in place of TextExpression at the beginning of the score. The lyrics were not tied to notes in that case, or the first note. We tried implementing a way to filter out using the length of TextExpression, assuming that if the lyrics are indeed included as part of it, the length of TextExpression would be more than say a specific threshold (maybe 30), however, many times the TextExpression consists of some details of the score like the composer etc. Hence, a basic thresholding based criteria does not provide sufficient information to filter based on TextExpression. We decided to go for the lyrics approach thereafter. We can add more musically informed criteria to filter out more renditions, however, since the original scores from CPDL in the data are not very consistent, it needs more thorough investigation to add more criteria. 

In general, the synthesizer operates satisfactorily, both in tuning and linguistics, but its performance depends on the suitability of the accompanying score. Consequently, an accurate OCR conversion from .pdf to .mscz files and error-free original scores are essential prerequisites for the synthesizer to function effectively. This requirement poses a significant challenge for a substantial portion of the available scores.

Furthermore, we identify instances in which the audio, while initially maintaining correct tuning, experiences distortion as it progresses towards the end of the note. Our observations suggest that this phenomenon is more prevalent in long-duration sounds, possibly attributed to tempo variations; however, it's important to note that this pattern is not consistently observed in all such cases. Similarly, the punctuation marks used to delineate phrases or segments within phrases are not consistently replicated, including commas, semicolons, periods, or the inherent speech cadences that we often internalize and consider as given. When it comes to sung phrases, it creates the impression of a vocalist who can sustain extended notes without taking a breath, resulting in an artificial quality in the synthesis.

Correct diction for different languages still remains a challenge, and the following languages can be listed, in an order from highest to lowest precision achieved upto this point: Latin (we include Spanish pieces here due to common origin of the languages), English, German. The pieces analyzed in Latin are quite precise. The few pieces in English that were erroneous had mainly to do with poor transcriptions of the parts or a mispronunciation. For the case of German language, there is still a need to achieve greater precision in the pronunciation of some phonemes, particularly diphthongs, while consonants - or combinations are well achieved.

Through our initial investigation, we find that a large number of expressions are not replicated in the same way as one would expect. For example, the expressions dynamics \cite{narang:22:dafx} relies primarily on volume changes while a real singer would consider additional aspects like changes in timbre or harmonies to add the effect of dynamics.  
 Hence, synthesis would need continued efforts from the side of musicians and technologists combined to aid in improved rendition quality.

\section{Future Work}
We believe the dataset developed is quite useful, and the methodology employed to be adaptable for extending the dataset in the future. Potential future directions entail  testing it in the context of MIR research tasks, and identifying any limitations that may lead to further dataset improvements. Additionally, there is room for enhancement of the quality of musicXML scores obtained from CPDL. Further enhancements may include development of the synthesizer capable of accurately modelling the expressive nuances present in the score, and creating a mix using the synthesized voices that could be representative of a real choral mix.

\bibliography{bibliography.bib}
\end{document}